\documentstyle[11pt,amssymb]{article}

\def\dalemb#1#2{{\vbox{\hrule height .#2pt
        \hbox{\vrule width.#2pt height#1pt \kern#1pt
                \vrule width.#2pt}
        \hrule height.#2pt}}}
\def\square{\mathord{\dalemb{6.8}{7}\hbox{\hskip1pt}}}

\def\cF{{\cal F}}
\def\cA{{\cal A}}

\def\0{{\sst{(0)}}}
\def\1{{\sst{(1)}}}
\def\2{{\sst{(2)}}}
\def\3{{\sst{(3)}}}
\def\4{{\sst{(4)}}}
\def\5{{\sst{(5)}}}
\def\6{{\sst{(6)}}}
\def\7{{\sst{(7)}}}
\def\8{{\sst{(8)}}}

\def\Z{\rlap{\sf Z}\mkern3mu{\sf Z}}
\def\R{\rlap{\rm I}\mkern3mu{\rm R}}

\def\ep{\epsilon}
\def\td{\tilde}
\def\wtd{\widetilde}

\let\a=\alpha

\def\nn{\nonumber} \def\bd{\begin{document}} \def\ed{\end{document}}
\def\ds{\documentstyle} \let\fr=\frac \let\bl=\bigl \let\br=\bigr
\let\Br=\Bigr \let\Bl=\Bigl 
\let\bm=\bibitem
\let\na=\nabla
\let\pa=\partial \let\ov=\overline 
\newcommand{\be}{\begin{equation}} 
\newcommand{\ee}{\end{equation}} 
\def\ba{\begin{array}}
\def\ea{\end{array}}
\def\ft#1#2{{\textstyle{{\scriptstyle #1}\over {\scriptstyle #2}}}}
\def\fft#1#2{{#1 \over #2}}
\def\del{\partial}
\def\sst#1{{\scriptscriptstyle #1}}
\def\oneone{\rlap 1\mkern4mu{\rm l}}
\def\ie{{\it i.e.\ }}
\def\via{{\it via}}
\def\semi{{\ltimes}}
\def\str{{\rm str}}
\def\jm{{\rm j}}
\def\im{{\rm i}}
\def\bOmega{{{\bar\Omega}}}
\def\Qn{{{Q_{\sst{\rm N}}}}}
\def\tX{{{\wtd X}}}

\def\mapright#1{\smash{\mathop{-\!\!\!-\!\!\!-\!\!\!-\!\!\!-\!\!\!
             \longrightarrow}\limits^{#1}}}
\def\maprightt#1#2{\smash{\mathop{-\!\!\!-\!\!\!-\!\!\!-\!\!\!-\!\!\!
             \longrightarrow}\limits^{#1}_{#2}}}

\newcommand{\ho}[1]{$\, ^{#1}$}
\newcommand{\hoch}[1]{$\, ^{#1}$}
\newcommand{\bea}{\begin{eqnarray}} 
\newcommand{\eea}{\end{eqnarray}} 
\newcommand{\ra}{\rightarrow}
\newcommand{\lra}{\longrightarrow}
\newcommand{\Lra}{\Leftrightarrow}
\newcommand{\ap}{\alpha^\prime}
\newcommand{\bp}{\tilde \beta^\prime}
\newcommand{\tr}{{\rm tr} }
\newcommand{\Tr}{{\rm Tr} }

\newcommand{\NP}{Nucl. Phys. }
\newcommand{\tamphys}{\it Center for Theoretical Physics\\
Texas A\&M University, College Station, Texas 77843}
\newcommand{\umich}{\it Department of Physics\\
University of Michigan, Ann Arbor, Michigan 48109}
\newcommand{\upenn}{\it Department of Physics and Astronomy\\
University of Pennsylvania, Philadelphia, Pennsylvania 19104}
\newcommand{\SISSA}{\it  SISSA-ISAS and INFN, Sezione di Trieste\\
Via Beirut 2-4, I-34013, Trieste, Italy}

\newcommand{\auth}{M. Cveti\v{c}\hoch{\dagger1}, 
H. L\"u\hoch{\star2} and C.N. Pope\hoch{\ddagger3}}

\begin{document}
\begin{flushright}
\hfill{CTP TAMU-28/00}\ \ \ {UPR-903-T}\ \ \
{UM-TH-00-23}\ \ \
{SISSA-Ref.88/2000/EP}\ \ \ 
{hep-th/0009183}\\
\hfill{September, 2000}\\
\end{flushright}


\begin{center}
{ \large {\bf Brane-world Kaluza-Klein Reductions\\
and Branes on the Brane }}

\vspace{10pt}
\auth

\vspace{10pt}
{\hoch{\dagger}\upenn}

\vspace{10pt}
{\hoch{\dagger}\hoch{\ddagger}\SISSA}

\vspace{10pt}
{\hoch{\star}\umich}

\vspace{10pt}
{\hoch{\ddagger}\tamphys}

\vspace{20pt}

\underline{ABSTRACT}
\end{center}

      We present a systematic study of a new type of consistent
``Brane-world Kaluza-Klein Reduction,'' which describe fully
non-linear deformations of co-dimension one objects that arise as
solutions of a large class of gauged supergravity theories in diverse
dimensions, and whose world-volume theories are described by 
ungauged supergravities with one half of the original supersymmetry.
In addition, we provide oxidations of these Ans\"atze which are in
general related to sphere compactified higher dimensional string
theory or M-theory.  Within each class we also provide explicit
solutions of brane configurations localised on the world-brane.  We
show that at the Cauchy horizon (in the transverse dimension of the
consistently Kaluza-Klein reduced world-brane) there is a curvature
singularity for any configuration with a non-null Riemann curvature or
a non-vanishing Ricci scalar that lives in the world-brane.  Since the
massive Kaluza-Klein modes can be consistently decoupled, they cannot
participate in regulating these singularities.

{\vfill\leftline{}\vfill
\footnoterule
{\footnotesize \hoch{1} Research supported in part by DOE grant 
DE-FG02-95ER40893 and NATO grant 976951. \vskip -12pt} \vskip 14pt
{\footnotesize \hoch{2} Research supported in full by DOE grant
DE-FG02-95ER40899 \vskip -12pt} \vskip 14pt
{\footnotesize  \hoch{3} Research supported in part by DOE 
grant DE-FG03-95ER40917.\vskip  -12pt}}

\pagebreak
\setcounter{page}{1}

\tableofcontents
\vfill\eject

\section{Introduction}

     The conventional way of extracting an effective lower-dimensional
theory from a higher-dimensional one is by performing a Kaluza-Klein
reduction in which the extra dimensions are wrapped up into a compact
space such as a torus or a sphere.  Provided that the scale size of
these internal dimensions is sufficiently small in relation to the
energy scale of excitations in the lower dimension, then the mass gap
separating the massless modes from the massive ones will be sufficient
to ensure that the internal dimensions are essentially unobservable, and
the world will appear to be effectively lower dimensional.

    If an extra dimension were non-compact then seen from the
lower-dimensional viewpoint there would usually be a continuum of modes,
with masses extending down to zero.  One would normally expect that this
would mean that the observable world would be the higher-dimensional
one, and that one could not usefully describe it in terms of a
lower-dimensional viewpoint.\footnote{We cannot usefully view our
four-dimensional spacetime as being effectively three-dimensional simply
by shutting our eyes to the existence of the $z$ axis!}  However, it has
been shown that under suitable circumstances it may be that the
continuous mass eigenvalues for the massive lower-dimensional metric
perturbations are distributed in such a way that the effects of the
nearly-massless modes is suppressed, implying that the world does in
fact appear to be lower-dimensional, with only small modifications to
the gravitational forced law appropriate to the lower dimension
\cite{rs2}.  In its original form this Randall-Sundrum II scenario is
realised by starting from pure gravity with a negative cosmological
constant in five dimensions, and patching together two segments of
AdS$_5$.  In horospherical coordinates one has
\be
d\hat s_5^2 = e^{-2k\, |z|}\, \eta_{\mu\nu}\, dx^\mu\, dx^\nu + dz^2\,,
\label{rsmetric}
\ee
where the 3-brane is located at $z=0$.\footnote{For a review on global
and local space-time structure of co-dimension one objects see
\cite{csrev}.}It was found that gravity is effectively localised on the
3-brane corresponding to the join between the two segments of AdS$_5$
\cite{rs2}.  Specifically, it was shown that the metric fluctuations
around the flat Minkowski spacetime of the 3-brane are localised near
the brane.

     More generally, if the flat Minkowski metric on the 3-brane is
replaced by any Ricci-flat 4-metric the five-dimensional metric will
still, in the bulk, satisfy the Einstein equations with a negative
cosmological constant.   In other words, one can view
\be
d\hat s_5^2 = e^{-2k\, |z|}\, ds_4^2 + dz^2\label{bwkk}
\ee
as a Kaluza-Klein reduction Ansatz that gives a consistent embedding
of four-dimensional pure Einstein gravity in five-dimensional Einstein
gravity with a negative cosmological constant.  In fact the construction
could be extended to give an embedding of four-dimensional $N=1$
ungauged supergravity in five dimensions, by starting from $N=2$ (\ie
minimal) gauged supergravity in $D=5$.  Note, however, that the
bosonic sector in $D=4$ would still only comprise the metric, and there
would be no Maxwell field that could support charged
Reissner-Nordstr\"om black holes.  In particular, it should be noted
that one cannot get a Maxwell field as a standard type of Kaluza-Klein
vector by writing $d\hat s_5^2 = e^{-2k\, |z|}\, ds_4^2 + (dz+
\cA_\1)^2$, since $\del/\del z$ is not a Killing vector.

     In a recent paper, it was shown that if one instead starts with
$N=4$ gauged supergravity in five dimensions, then it is possible to
construct a consistent Kaluza-Klein reduction Ansatz that gives an
embedding of four-dimensional ungauged $N=2$ supergravity on the 3-brane
\cite{bob}.  This is a new kind of dimensional reduction, which we shall
refer to as ``Brane-world Kaluza-Klein Reduction.''  It should be
emphasised that it is non-trivial that a {\it consistent} Kaluza-Klein
reduction of this sort is possible,\footnote{A consistent reduction is
one where all the higher-dimensional equations of motion are satisfied
provided that the lower-dimensional fields satisfy their equations of
motion.} and there is no obvious group-theoretic explanation for why it
should work.  Two further examples of consistent brane-world
Kaluza-Klein reductions were obtained in \cite{bob}, one describing the
embedding of six-dimensional ungauged chiral $N=(1,0)$ supergravity in
seven-dimensional $SU(2)$-gauged $N=2$ supergravity, and the other
describing the embedding of five-dimensional ungauged $N=2$ supergravity
in six-dimensional $SU(2)$-gauged $N=2$ supergravity.  More generally,
it was conjectured that it should be possible to find a consistent
brane-world Kaluza-Klein reduction from any gauged supergravity in $D$
dimensions to an ungauged supergravity with half the supersymmetry in
$(D-1)$ dimensions \cite{bob}.

   The purpose of this paper is to provide a systematic construction of
consistent brane-world Kaluza-Klein reductions for gauged supergravity
theories (with maximal supersymmetry) in diverse dimensions, thus in
general leading to the ungauged supergravities with one-half of the
original (maximal) gauged supersymmetry.  In addition, the studied
examples provide compactifications on both AdS and dilatonic
co-dimension one objects.  In the first of these examples, in section 2,
we show that five-dimensional maximal ($N=8$) $SO(6)$-gauged
supergravity admits a consistent brane-world reduction to
four-dimensional ungauged $N=4$ supergravity.  Next, in section 3, we
show that massive type IIA supergravity admits a consistent brane-world
reduction to nine-dimensional ungauged $N=1$ supergravity.  Next, in
section 4 we show that eight-dimensional maximal $SU(2)$-gauged
supergravity admits a consistent brane-world reduction to
seven-dimensional ungauged $N=2$ supergravity.  Then, in section 5, we
show that seven-dimensional maximal $SO(5)$-gauged supergravity admits a
consistent brane-world reduction to seven-dimensional ungauged $N=(2,0)$
chiral supergravity.

    All the brane-world reductions that were constructed in
\cite{bob}, and the brane-world reductions of five-dimensional maximal
$SO(6)$-gauged supergravity and seven-dimensional maximal
$SO(5)$-gauged supergravity in this paper, are examples where the
higher-dimensional theory admits an anti-de Sitter vacuum solution.
By contrast, massive type IIA supergravity and the eight-dimensional
$SU(2)$-gauged supergravity that we also consider in this paper do not
admit anti-de Sitter solutions, but instead they have dilatonic domain
walls as their most symmetric ``vacuum'' solutions.  In all the cases,
the brane-world Kaluza-Klein reductions can be thought of as fully
non-linear descriptions of deformations around the anti-de Sitter or
domain-wall background, in which the $(D-1)$-dimensional Minkowski
metric on the $(D-2)$-brane in the $D$-dimensional AdS or domain-wall
vacuum is allowed to become arbitrary, along with the other necessary
fields that complete the $(D-1)$-dimensional ungauged supergravity
multiplet.\footnote{We should emphasise that as with any fully
non-linear Kaluza-Klein Ansatz, the reduction is not pinned to any
specific solution.  Although it may sometimes be convenient to think
of the AdS or domain-wall solution as playing a preferred r\^ole, it
is really just one out of an infinity of solutions of the reduced theory.}
   
    A brane-world type of Kaluza-Klein reduction can also be performed
in those cases where the $p$-brane cannot trap gravity.  A
classification of domain walls that can and cannot trap gravity was
given in \cite{youm,clpdw}. In these cases, gravity can arise by placing
the $p$-brane on orbifold points \cite{losw} \`a la Ho\v rava-Witten
\cite{hw}.

      At the level of the supergravity theory, the requirement of the
consistency of the brane-world Kaluza-Klein reduction does not
discriminate between whether or not the brane is capable of trapping
gravity.  This is analogous to the situation for a standard
Kaluza-Klein reduction on $S^1$; at the level of the massless modes,
which are the only ones retained in the consistent truncation, one
cannot distinguish between an extra dimension that is a circle or an
infinite real line.  In particular, we shall usually write the
brane-world reduction Ansatz, as in (\ref{bwkk}), with an
absolute-value sign for the coordinate $z$, \ie insisting on a $Z_2$
symmetric co-dimension one Ansatz for the transverse dimension. Thus,
there is an actual delta function source needed at $z=0$, whose origin
lies outside the supergravity Lagrangian description. However, from
the mathematical point of view we could perfectly well write the
Ansatz without the absolute-value sign, thus describing the bulk
solution only, which would still correspond to a consistent reduction.
In fact now the Ansatz will satisfy the equations everywhere, without
the need for any external delta-function sources. (However, one would
now lose the brane-world interpretation (at $z=0$) of the reduced
theory.)

   We conclude this Introduction with a table that summarises the
principal results that we obtain in this paper, and those of
\cite{bob}.

\bigskip\bigskip
\centerline{
\begin{tabular}{|c|c|c|}\hline
$D$   & $D$-dimensional Theory & $(D-1)$-dimensional Theory from \\
      &                        &Brane-world Reduction\\ \hline\hline
10 & Massive IIA & $D=9$, $N=1$  \\ \hline
8 & $SU(2)$-gauged $N=2$  & $D=7$, $N=2$  \\ \hline
7 & $SO(5)$-gauged $N=4$  & $D=6$, $N=(2,0)$  \\ \hline
6 & $SU(2)$-gauged $N=2$ & $D=5$, $N=2$  \\ \hline
5 & $SO(6)$ gauged $N=8$ & $D=4$, $N=4$  \\ \hline
\end{tabular}}
\bigskip

\noindent{\bf Table 1:} The ungauged supergravities in $(D-1)$
dimensions obtained by brane-world \phantom{xxxxxxxxx} 
Kaluza-Klein reductions.
\bigskip

    In addition, we shall consider $S^1$ reductions of the $D=8$ and $D=7$
gauged supergravities.  The former provides a brane-world reduction from
$D=7$ that gives the non-chiral $N=(1,1)$ ungauged supergravity in
$D=6$, while the latter provides a brane-world reduction from $D=6$ that
gives the $N=4$ ungauged theory in $D=5$.  The brane-world reduction of
the $D=6$ $SU(2)$-gauged supergravity was obtained in \cite{bob}, as
were the brane-world reductions of the $SU(2)$-gauged $N=2$
seven-dimensional supergravity, and the $SU(2)\times U(1)$ gauged $N=4$
five-dimensional supergravity.  These two cases are contained within
reductions with larger supersymmetries that we consider here.  It should
be noted that an intrinsic feature of brane-world Kaluza-Klein
reductions is that the reduced theory never has more than half the
maximal supersymmetry that is allowed in that dimension.  This is
associated with the fact that there is always a halving of supersymmetry
on the brane solution of the higher-dimensional gauged or massive
supergravity.

\section{Four-dimensional $N=4$ supergravity from maximal \\five-dimensional 
gauged supergravity}

\subsection{Direct reduction from type IIB supergravity}

   In section 2.2 below, we shall obtain the brane-world embedding of
four-dimensional ungauged $N=4$ supergravity in five-dimensional maximal
gauged supergravity, thus providing the brane-world Kaluza-Klein
compactification in D=4 with the maximal allowed ungauged supersymmetry.
However, since this five-dimensional theory is rather complicated, we
shall begin in the current section by constructing the brane-world
embedding of the four-dimensional $N=4$ theory directly in
ten-dimensional type IIB supergravity.  This exploits the fact that the
five-dimensional gauged theory can itself be obtained via an $S^5$
reduction from $D=10$.  Having done this, we shall then be in a position
to re-express our results in terms of a brane-world reduction from $D=5$
to $D=4$.  From the five-dimensional viewpoint the fields that we use
are the metric, the dilaton and axion (which are singlets under the $SO(6)$
gauge group), and the two sets of six 2-form potentials.  Thus in $D=5$
the 15 Yang-Mills gauge fields and the $10+\overline{10} + 20'$ of
scalars are set to zero.

    The bosonic equations of motion of type IIB supergravity can be
derived from the Lagrangian
\bea
{\cal L}^{IIB}_{10} &=& \hat R\, {\hat *\oneone} - \ft12
{\hat *d\hat{\phi}}\wedge d\hat{\phi} - \ft12
e^{2\hat{\phi}}\, 
{\hat *d\hat{\chi}}\wedge d\hat{\chi} - \ft14 {\hat *\hat F_\5}
\wedge \hat F_\5 \nn\\
& & -\ft12 e^{-\hat \phi} \, {\hat * \hat{F}^{2}_\3}\wedge\hat{F}^{2}_\3 -
\ft12 e^{\hat\phi}\, {\hat *\hat{F}^{1}_\3}\wedge \hat{F}^{1}_\3 - 
\ft12 \hat A_\4\wedge d\hat{A}_\2^{1}\wedge d\hat{A}_\2^{2}\ ,\label{d10lag}
\eea
where $\hat F^2_\3 = d\hat A^2_\2,\, \hat F^1_\3=d\hat
A^1_\2 - \hat{\chi}\, d\hat{A}^2_\2,\, \hat{F}_\5=d\hat{A}_\4 -
\ft12 \hat{A}_\2^1\wedge d\hat{A}_\2^2 + \ft12
\hat{A}_\2^2\wedge d\hat{A}_\2^1$, and we use hats to denote
ten-dimensional fields and the ten-dimensional Hodge dual $\hat
*$. The equations of motion following from the Lagrangian, together
with the self-duality condition, are
\bea
\hat R_{MN} &=& \ft12\pa_M\hat{\phi}\, \pa_N\hat{\phi} + \ft12 e^{2
\hat\phi}\, \pa_M\hat{\chi}\, \pa_N\hat{\chi} + \ft1{96} \hat F^2_{MN}
+ \ft14 e^{\hat\phi}\, \Big((\hat{F}^{1}_\3)^2_{MN} -
\ft1{12}(\hat{F}^1_\3)^2 \hat{g}_{MN}\Big)\nn\\
& & + \ft14 e^{-\hat\phi}\,
\Big((\hat{F}^{2}_\3)^2_{MN} - \ft1{12}
(\hat{F}^2_\3)^2\hat{g}_{MN}\Big)\,,\nn\\
d{\hat * d\hat{\phi}} &=& - e^{2\hat{\phi}}\, {\hat * d\hat\chi}\wedge
d\hat{\chi} - \ft12 e^{\hat{\phi}}\, {\hat *\hat{F}_\3^1}\wedge
\hat{F}_\3^1 + \ft12 e^{-\hat{\phi}}\, {\hat * \hat{F}_\3^2}\wedge
\hat{F}_\3^2\,,\nn\\
d\Big(e^{2\hat{\phi}}\, {\hat * d\hat\chi}\Big) &=&
e^{\hat{\phi}}\, {\hat * \hat{F}_\3^1}\wedge \hat{F}_\3^2\,,\nn\\
d\Big(e^{\hat{\phi}}\, {\hat * \hat{F}_\3^1}\Big) &=& \hat{F}_\5\wedge
\hat{F}_\3^2\,,\qquad  
d\Big(e^{-\hat{\phi}}\, {\hat *\hat{F}_\3^2} -
\hat{\chi}\, e^{\hat{\phi}}\, {\hat *\hat{F}_\3^1}\Big) = -
\hat{F}_\5\wedge ( \hat{F}_\3^1 + \hat{\chi}\, \hat{F}_\3^2)\,,\nn\\
d({\hat *\hat{F}_\5}) &=& - \hat{F}_\3^1\wedge
\hat{F}_\3^2, \qquad  \hat{F}_\5 = {\hat *\hat{F}_\5}\,.\label{2beom}
\eea

    The ungauged four-dimensional $N=4$ supergravity that we are seeking
to embed in type IIB supergravity is described by the following
Lagrangian:\footnote{Actually this Lagrangian corresponds to a special
truncation of toroidally compactified heterotic string theory where the
gauge fields of the original heterotic string are turned off and the
momentum and winding modes of the NS-NS sector are
identified, thus freezing the internal metric and antisymmetric two-form
fields of the six-torus.}
\be {\cal L}_4 = R\, {*\oneone} - \ft12 {*d\phi}\wedge d\phi - 
\ft12 e^{2\phi}\,
{*d\chi}\wedge d\chi - \ft12 e^{-\phi}\, {*F_\2^i}\wedge F_\2^i  
 - \ft12\chi\, F_\2^i\wedge F_\2^i\,,\label{d4n4lag}
\ee
where $1\le i\le 6$, and $F_\2^i=dA_\1^i$.  

    We find that the following is a consistent reduction Ansatz that
gives the embedding of the four-dimensional $N=4$ theory in type IIB
supergravity:
\bea
d\hat s_{10}^2 &=& e^{-2k\, |z|}\, ds_4^2 + dz^2 + g^{-2}\,
d\Omega_5^2\,,\nn\\
\hat F_\5 &=& 4g^{-4}\, \Omega_\5 + 4g\, e^{-4k\, |z|}\, \ep_\4\wedge
dz\,,\nn\\
\hat A_\2^1 &=& \ft1{\sqrt2}\, g^{-1}\, e^{-k\, |z|}\, \mu_i\,
(e^{-\phi}\, {*F_\2^i} + \chi\, F_\2^i)\,,\nn\\
\hat A_\2^2 &=& \ft1{\sqrt2}\, g^{-1}\, e^{-k\, |z|}\, \mu_i\,
F_\2^i\,,\nn\\
\hat\phi &=&\phi\,,\qquad \hat\chi = \chi\,,\label{10to4ans}
\eea
where $d\Omega_5^2$ is the metric on the unit 5-sphere, which we can write
in terms of six coordinates $\mu_i$ that are subject to the constraint
$\mu_i\, \mu_i=1$, as $d\Omega_5^2 = d\mu_i\, d\mu_i$.  
The 5-form $\Omega_\5$ is the volume form of the
metric $d\Omega_5^2$, and $\ep_\4$ is the volume form of the metric
$ds_4^2$.  Note that $\Omega_\5$ can be written as
\be
\Omega_\5 = \ft1{5!}\, \ep_{i j_1\cdots j_5}\, \mu_i\,
d\mu_{j_1}\wedge \cdots \wedge d\mu_{j_5}\,.
\ee
The constant $k$ (which we take to be positive) is related to the
gauge-coupling constant $g$ of the five-dimensional theory by
$k^2=g^2$.  In fact, to be precise, we must have
\be
g=  \left\{\matrix{ k\,, & z>0\,, \cr
                    -k\,, & z<0\,. } \right. \label{kgrel5}
\ee

    Substituting the Ansatz (\ref{10to4ans}) into the equations of
motion of type IIB supergravity (\ref{2beom}), we find that they are all
exactly satisfied if and only if the four-dimensional fields $ds_4^2$,
$\phi$, $\chi$ and $F_\2^i$ satisfy the equations of motion of ungauged
$N=4$ supergravity, which can be derived from (\ref{d4n4lag}).  Note in
particular that the six abelian gauge fields $F_\2^i$ satisfy the
equations of motion
\be
d(e^{-\phi}\, {*F_\2^i} + \chi\, F_\2^i)=0\,.\label{d4f2eom}
\ee
The following results are useful for verifying the consistency of the
reduction Ansatz.  Firstly, we have from (\ref{10to4ans}) that
\bea
\hat F_\3^1 &=&\ft1{\sqrt2}\, g^{-1}\, 
 e^{-k\, |z|}\, e^{-\phi}\, {*F_\2^i}\wedge (d\mu_i -g\,
\mu_i\, dz)\,,\nn\\
\hat F_\3^2 &=&\ft1{\sqrt2}\, g^{-1}\, 
 e^{-k\, |z|}\, F_\2^i\wedge (d\mu_i - g\, \mu_i\, dz)\,.
\eea
(Here, we have for convenience of presentation already made use of the
fact that the $F_\2^i$ satisfy the Bianchi identities $dF_\2^i=0$ and
the field equations (\ref{d4f2eom}); they can, of course, be {\it
derived} by substituting the Ansatz into the ten-dimensional equations
of motion.)  Next, we can write the ten-dimensional Hodge duals of
these field strengths as follows:
\bea
{\hat *\hat F_\3^1} &=& \ft1{\sqrt2}\, g^{-5}\, e^{-k\, |z|}\,
e^{-\phi}\, F_\2^i\wedge (\mu_i\, \Omega_\5 - g\, Z_i\wedge dz)\,,\nn\\
{\hat *\hat F_\3^2} &=& -\ft1{\sqrt2}\, g^{-5}\, e^{-k\, |z|}\,
{*F_\2^i}\wedge (\mu_i\, \Omega_\5 - g\, Z_i\wedge dz)\,,
\eea
where the 4-form $Z_i$ is defined by
\be
Z_i \equiv \ft1{4!}\, \ep_{ij k_1\cdots k_4}\, \mu_j\,
d\mu_{k_1}\wedge \cdots \wedge d\mu_{k_4}\,.
\ee
This 4-form is the Hodge dual of $d\mu_i$ in the unit 5-sphere metric, 
${*_5d\mu_i} = -Z_i$.  Note that $Z_i$ has the following
properties:
\be
d\mu_i\wedge Z_j = -(\delta_{ij} -\mu_i\, \mu_j)\, \Omega_\5\,,\qquad
dZ_i = 5\mu_i\, \Omega_\5\,.
\ee
It is now straightforward to verify that all the type IIB
ten-dimensional equations of motion consistently yield the equations of
motion of four-dimensional ungauged $N=4$ supergravity; in particular,
all the dependence on the coordinates $z$ and $\mu_i$ consistently
matches in all the equations.

   It is worth noting that the $N=4$ gauged supergravity in the
four-dimensional world-volume of the D3-brane has an $SL(2,\Z)$
electric/magnetic S-duality, with the two scalars $(\phi,\chi)$
parameterising the $SL(2,\R)/O(2)$ coset.  It is easy to see from the
reduction Ansatz (\ref{10to4ans}) that this $SL(2,\Z)$ symmetry of the
theory in the world-volume of the D3-brane originates from the $SL(2,\Z)$ of
the original type IIB theory in $D=10$, which is not an
electric/magnetic duality.

    It is of interest to see how the brane-world embedding of the
four-dimensional $N=4$ supergravity that we have derived here reduces to
the $N=2$ supergravity embedding that was constructed in \cite{bob}.  In
the $N=2$ theory there is just one 2-form field strength $F_\2$, and the
dilaton $\phi$ and axion $\chi$ are absent.  It is easy to see that the
equations of motion for $\phi$ and $\chi$ in (\ref{d4n4lag}) imply that
in order to set $\phi=\chi=0$, we must have
\be
{*F_\2^i}\wedge F_\2^i=0\,,\qquad F_\2^i\wedge F_\2^i=0\,.
\ee
The minimal non-trivial way to satisfy these conditions is by taking all
but two of the six field strengths to vanish, and for the remaining
ones, say $F_\2^1$ and $F_\2^2$, to be related by $F_\2^2={*F_\2^1}$.
If we define $F_\2^2={*F_\2^1}=-F_\2/\sqrt2$, and at the same time we
parameterise the six coordinates $\mu_i$ that define the 5-sphere as
\be
\mu_1=\sin\xi\, \cos\tau\,,\qquad \mu_2 = -\sin\xi\, \sin\tau\,,
\qquad \mu_\a = \nu_\a\, \cos\xi\,,\qquad (\a=3,4,5,6)\,,
\ee
where $\nu_\a\, \nu_\a=1$, defining a unit 3-sphere, the Ans\"atze for
$d\hat s_{10}^2$, $\hat A_\2^1$ and $\hat A_\2^2$ in (\ref{10to4ans})
become
\bea
d\hat s_{10}^2 &=& e^{-2k\, |z|}\, ds_4^2 + dz^2 + d\xi^2 + \sin^2\xi\,
d\tau^2 + \cos^2\xi\, d\Omega_3^2\,,\nn\\
\hat A_\2^1 + \im\, \hat A_\2^2 &=& -\ft12 g^{-1}\, e^{-k\, |z|}\,
\sin\xi\, e^{-\im\, \tau}\, (F_\2- \im\, {*F_\2})\,.
\eea
This is precisely the form of the Ansatz found in \cite{bob} for the
consistent brane-world embedding of four-dimensional $N=2$ supergravity.

\subsection{Ungauged $D=4$, $N=4$ from gauged $D=5$, $N=8$}

    In the previous subsection, we considered the reduction from type
IIB to the ungauged $N=4$ theory in $D=4$ directly, omitting the
intermediate description as a brane-world KK reduction of
five-dimensional $N=8$ gauged supergravity on account of the complexity
of the five-dimensional theory.  We can now in fact reinterpret our
results as a reduction of the maximal five-dimensional gauged theory.
However, in order to avoid the full complexity of this theory, we shall
work with a truncation of the full set of five-dimensional fields in
which just the metric, the dilaton $\phi$ and axion $\chi$, and the
$6+6$ of 2-form potentials are retained.  In other words, we set the 15
Yang-Mills $SO(6)$ gauge fields and the $10 + \overline{10} +20'$ of
scalars to zero.  It should be emphasised that this is in general an
{\it inconsistent} truncation of the five-dimensional theory.  However,
we can still work with it provided that we impose the necessary
algebraic constraints on the $6+6$ of 2-form potentials.  Of course
these constraints are precisely the ones that {\it are} satisfied by the
brane-world KK reduction Ansatz, now expressed simply as a reduction
from $D=5$ to $D=4$.

   We find that the above truncated 5-dimensional theory is obtained
from $D=10$ by making the following Ansatz for the type IIB fields:
\bea
d\hat s_{10}^2 &= &ds_5^2 + g^{-2}\,
d\Omega_5^2\,,\nn\\
\hat F_\5 &=& 4g^{-4}\, \Omega_\5 + 4g\, \ep_\5 \,,\nn\\
\hat A_\2^\a &=& \mu_i\, A_2^{i\a}\,,\nn\\
\hat\phi &=&\phi\,,\qquad \hat\chi = \chi\,,\label{10to5ans}
\eea
where $\a=1,2$.  We immediately find from the Bianchi identity for
$\hat F_\5$ that the following equations must be satisfied:
\be
A_\2^{i\a}\wedge A_\2^{j\beta}\, \ep_{\a\beta}=0\,,\qquad
dA_\2^{i\a}\wedge A_\2^{j\beta}\, \ep_{\a\beta}=0\,.\label{constraints}
\ee
These are the algebraic constraints alluded to above.  We must impose
them because we have set other fields of the maximal five-dimensional
theory to zero, which is in general in conflict with the equations of
motion of those fields.

   Substituting the Ansatz (\ref{10to5ans}) into the remaining
equations of motion of type IIB supergravity, and making use of the
constraints (\ref{constraints}), we find that they
consistently imply the following five-dimensional equations:
\bea
dA_\2^{i1}-\chi\, dA_\2^{i2}  &=& - g\, e^{-\phi}\,
{*A_\2^{i2}}\,,\nn\\
dA_\2^{i2} &=& g\, e^\phi\, {*(A_\2^{i1} -\chi\, A_\2^{i2})}\,,\nn\\
d(e^{2\phi}\, {*d\chi}) &=& -g^2\, e^\phi\,{* (A_\2^{i1} - \chi\,
A_\2^{i2})} \wedge A_\2^{i2}\,,\nn\\
d{*d\phi} &=& e^{2\phi}\, {*d\chi}\wedge d\chi + \ft12 g^2\, e^\phi\, 
{*(A_\2^{i1} -\chi\, A_\2^{i2})}\wedge (A_\2^{i1} -\chi\, A_\2^{i2})
\nn\\
&&-\ft12 g^2\, e^{-\phi}\, {*A_\2^{i2}}\wedge A_\2^{i2}\,,\nn\\
R_{\mu\nu} &=& \ft12 \del_\mu\phi\, \del_\nu\phi + \ft12 e^{2\phi}\,
\del_\mu\chi\, \del_\nu\chi \nn\\
&&+ \ft12 g^2\, \Big[ e^\phi\,(A_{\mu\rho}^{i1} -\chi\, A_{\mu\rho}^{i2})\,
(A_\nu^{i1\, \rho}-\chi\, A_\nu^{i2\, \rho} ) + e^{-\phi}\, 
A_{\mu\rho}^{i2}\, A_\nu^{i2\, \rho} \Big]\,.\label{d5eoms}
\eea
($A_\2^{i1}$ and $A_\2^{i2}$ denote $A_\2^{i\a}$ with $\a=1$ and $\a=2$
respectively.)  These equations, together with the constraints
(\ref{constraints}), are precisely equivalent to those of maximal
five-dimensional gauged supergravity, after setting the Yang-Mills
fields and the $10+\overline{10} + 20'$ of scalars to zero.  The $6+6$
of 2-form fields $A_\2^{i\a}$ satisfy first-order equations of motion,
known as ``odd-dimensional self-duality equations.''  These, together
with the constraint equations (\ref{constraints}), imply that the trace
of the 2-form contributions in the energy-momentum tensor vanishes;
$e^\phi\,(A_{\mu\nu}^{i1} -\chi\, A_{\mu\nu}^{i2})(A^{i1\,
\mu\nu}-\chi\, A^{i2\, \mu\nu} ) + e^{-\phi}\, A_{\mu\nu}^{i2}\, A^{i2\,
\mu\nu} =0$.  Note that the imposition of the constraints
(\ref{constraints}) is sufficient to ensure that {\it all} the type IIB
equations of motion are consistently satisfied by the Ansatz
(\ref{10to5ans}), including the internal and mixed components of the
Einstein equations.

  It is useful to observe that the equations of motion (\ref{d5eoms})
can be derived from the Lagrangian
\bea
{\cal L}_5 &=& R\, {*\oneone} -\ft12 {*d\phi}\wedge d\phi -\ft12
e^{2\phi}\, {*d\chi}\wedge d\chi -\ft12 g^2\, e^\phi\, {*(A_\2^{i1}
 -\chi\, A_\2^{i2})}\wedge  (A_\2^{i1} -\chi\, A_\2^{i2})\nn\\
&& - \ft12 g^2\, e^{-\phi}\,
{*A_\2^{i2}}\wedge A_\2^{i2} - g\, dA_\2^{i1} \wedge A_\2^{i2}
+ 12g^2\, {*\oneone} \,.
\eea

   Finally, we note that the brane-world Kaluza-Klein reduction of the
previous subsection, now expressed as a reduction from $D=5$ to $D=4$,
is given by
\bea
ds_5^2 &=& e^{-2k\, |z|}\, ds_4^2 + dz^2\,,\nn\\
A_\2^{i1} &=& \ft1{\sqrt2}\, g^{-1}\, e^{-k\, |z|}\, (e^{-\phi}\,
{*F_\2^i} + \chi\, F_\2^i)\,,\qquad
A_\2^{i2} = \ft1{\sqrt2}\, g^{-1}\, e^{-k\, |z|}\, F_\2^i\,,
\eea
with $\phi$ and $\chi$ just reducing directly.  One can easily verify
that this reduction Ansatz is indeed compatible with the constraints
(\ref{constraints}).

\subsection {Branes on the D3-brane}

       One can construct electric and magnetic black holes, strings and
instantons in $D=4$, $N=4$ supergravity.  They become branes on the
D3-brane (in the near-horizon region) when they are lifted back to $D=10$.  
We analyse these solutions in this section.

\bigskip\bigskip
\noindent{\it Case 1: $SL(2,\Z)$ dyonic black holes on D3-brane:}
\bigskip

      We can use one of the six 2-form field strengths to construct an
electric or magnetic black hole.  As a concrete example, let us consider
an electric black hole supported by the field strength
$F_\2^1$. Once the solution is lifted back to $D=10$, it
becomes
\bea
d\hat s_{10}^2 &=& e^{-2k|z|} \Big[-H^{-1}\, dt^2 + H\, (dr^2 + r^2
d\Omega_2^2)\Big] + dz^2 + g^{-2}\, d\Omega_5^2\,,\nn\\
\hat F_\5 &=& 4g^{-4}\, \Omega_5 + 4g\, e^{-4k|z|}\, r^2 H\, dt\wedge
dr\wedge \Omega_2\,,\nn\\
\hat A^1_\2 &=& \fft{Q}{\sqrt2}\,g^{-1}\, e^{-k|z|}\, \mu_1\, \Omega_2\,,
\nn\\
\hat A^2_\2 &=& \ft1{\sqrt2}\, g^{-1}\, e^{-k|z|}\, \mu_1\, dt\wedge
dH^{-1}\,,\nn\\
e^{\phi} &=& H\,,\qquad H=1 + \fft{Q}{r}\,,\label{ebond3}
\eea
where $\mu_1$ is one of the coordinates $\mu_i$ for $S^5$ appearing in
the Ansatz (\ref{10to5ans}), corresponding to our choice to consider a
black hole supported by $F_\2^1$.  Starting with the electric black
hole, we can then apply the $SL(2,\R)$ symmetry to get a multiplet of
dyonic black holes, where the electric and magnetic charges are carried
by the same 2-form field strength.  The metric of this dyonic solution
remains unchanged, but the charge configuration alters.

\bigskip\bigskip
\noindent{\it Case 2: Threshold dyonic black holes on D3-brane:}
\bigskip

     In $D=4$, $N=4$ supergravity, one can also construct a multi-charge
black hole solution, where the electric charge is carried by one 2-form
field strength, say $F_\2^1$, and the magnetic charge is carried by
another, say $F_\2^2$ \cite{kallosh}. (Note that
the generating technique, as employed, for example, for the four-charge
solution \cite{cy} of toroidally compactified heterotic string, may allow
for a construction of more general dyonic black holes with all the
$U(1)$ charges turned on.) Lifting this solution back to $D=10$, it
becomes
\bea
d\hat s_{10}^2 &=& e^{-2k|z|} \Big[-(H_1H_2)^{-1}\, dt^2 + 
  H_1H_2\, (dr^2 + r^2
d\Omega_2^2)\Big] + dz^2 + g^{-2}\, d\Omega_5^2\,,\nn\\
\hat F_\5 &=& 4g^{-4}\, \Omega_5 + 4g\, e^{-4k|z|}\, r^2 (H_1H_2)\, dt\wedge
dr\wedge d\Omega_2\,,\nn\\
\hat A^1_\2 &=& \fft{1}{\sqrt2}\,g^{-1}\, e^{-k|z|}\, (Q_e\, \mu_1\,
\Omega_2+ \mu_2\,  dt\wedge dH_2^{-1})\,,
\nn\\
\hat A^2_\2 &=& \ft1{\sqrt2}\, g^{-1}\, e^{-k|z|}\, (\mu_1\, dt\wedge
dH_1^{-1}+ \mu_2\,  Q_m\, \Omega_2)\,,\nn\\
e^{\phi} &=& \fft{H_1}{H_2}\,,\qquad 
H_1=1 + \fft{Q_e}{r}\,,\qquad H_2=1 + \fft{Q_m}{r}\,.
\label{embond3}
\eea

\bigskip\bigskip
\noindent{\it Case 3: String on D3-brane:}
\bigskip

      A magnetic string (four-dimensional domain wall), supported by the
axion, exists in the four-dimensional supergravity theory.  Lifting this
solution back to $D=10$, we have a string living on the D3-brane:
\bea
d\hat s_{10}^2 &=& e^{-2k|z|}\Big [-dt^2 + dx^2 + H(dr^2 + r^2d\theta^2)
\Big] + dz^2 + g^{-2} d\Omega_5^2\,,\nn\\
\hat F_\5 &=& 4g^{-4}\, \Omega_\5 + 4g\, e^{-4k|z|}\, rH\, 
dt\wedge dx \wedge dr \wedge d\theta\,,\nn\\
e^{\hat\phi}&=&H^{-1}\,,\qquad \chi=Q\, \theta\,,\qquad
H=1 + Q\, \log r\,.\label{stringond3}
\eea
This solution is a non-standard intersection of a D3-brane and D7-brane,
where there is no overall transverse space.  It should be distinguished
from the solution describing a D3-brane in the D7-brane, which has a
two-dimensional overall transverse space.

\bigskip\bigskip
\noindent{\it Case 4: Instanton on D3-brane:}
\bigskip

        The axion in the $D=4$ theory also supports a BPS instanton
solution when the theory is Euclideanised.  The axion $\chi$ becomes
imaginary under this procedure, the metric $ds_4^2$ becomes purely flat
Euclidean space, and $\hat F_\5$ becomes complex, since in ten Euclidean
dimensions a real 5-form cannot be self-dual.

\section{$N=1$ supergravity in $D=9$ from massive type IIA}
\subsection{D8-brane in massive type IIA theory}

     The highest dimensional D-brane that can be found in any
supergravity theory is the D8-brane in massive type IIA supergravity.
This theory was constructed in \cite{romans10}, but in a formulation
where there is not a straightforward massless limit to ordinary type IIA
supergravity.  However, it is simply a matter of performing a field
redefinition to resolve this problem \cite{78duality}.  The Lagrangian
for the bosonic sector of the massive type IIA supergravity can then be
written as the following differential form \cite{tension}
\bea
{\cal L}_{10} &=& \hat R\, {\hat *\oneone} -
\ft12 {\hat *d\hat \phi}\wedge d\hat \phi
 - \ft12 e^{\fft32\hat\phi}\, {\hat *\hat F_\2}\wedge \hat F_\2 -
\ft12 e^{-\hat \phi}\, {\hat *\hat F_\3}\wedge \hat F_\3-
\ft12 e^{\fft12\hat\phi}\, {\hat *\hat F_\4}\wedge \hat F_\4 \nn\\
&&\!\!
 -\ft12 d\hat A_\3\wedge d\hat A_\3 \wedge \hat A_\2 - \ft16 m\,
d\hat A_\3 \wedge (\hat A_\2)^3
 -\ft1{40} m^2\, (\hat A_\2)^5 -\ft12 m^2\, e^{\fft52\hat \phi}\,
{\hat *\oneone}\,,\label{romans1}
\eea
where the field strengths are given in terms of potentials by
\bea
\hat F_\2 &=& d\hat A_\1 + m\, \hat A_\2\ ,\qquad \hat F_\3 =
d\hat A_\2\,,\nn\\
\hat F_\4 &=& d\hat A_\3 + \hat A_\1\wedge d\hat A_\2 + \ft12 m\,
\hat A_\2\wedge \hat A_\2\,.\label{romfields}
\eea
The Bianchi identities for the field strengths are therefore
\be
d\hat F_\2 = m\, \hat F_\3\,,\qquad d\hat F_\3=0\,,\qquad 
d\hat F_\4 = \hat F_\2\wedge \hat F_\3\,,
\ee
and the field equations are
\bea
&&d\hat F_\6 = -\hat F_\3\wedge \hat
F_\4 \,, \qquad d \hat F_\8 = -\hat F_\3\wedge \hat F_\6\,,\nn\\
&&d\hat F_\7 = -\ft12 \hat F_\4\wedge\hat F_\4 - m\, \hat F_\8 - \hat
F_\2\wedge \hat F_\6\,,\\
&&d{\hat *d\hat \phi} = -\ft14\hat F_\4\wedge\hat F_\6 - \ft34 \hat
F_\2\wedge \hat F_\8 -\ft12\hat F_\3\wedge \hat F_\7 + \ft54 m^2\,
e^{\fft52\hat\phi}\, {\hat *\oneone}\,,\nn
\eea
where we have defined the dual field strengths
\be
\hat F_\6\equiv e^{\fft12\hat \phi}\, {\hat *\hat F_\4}\,,\qquad
\hat F_\8\equiv e^{\fft32\hat\phi}\, {\hat * \hat F_\2}\,,\qquad
\hat F_\7 \equiv e^{-\hat\phi}\, {\hat * \hat F_\3}\,.
\ee

       This massive type IIA theory supports a ``vacuum'' solution,
namely the D8-brane:
\be
d\hat s_{10}^2 = W^{\ft2{25}}\, dx^\mu dx_\mu + dz^2\,,\qquad
e^{\hat \phi} = W^{-\ft45}\,,\label{d8brane}
\ee
where the one-dimensional harmonic function is given by
\be
W=1 + k |z|\,,\qquad k^2 = \fft{625}{256}\, m^2\,.\label{wdef}
\ee
In fact the sign of $m$ must be opposite on opposite sides of the
domain wall:
\be
g=  \left\{\matrix{  \ft{16}{25} k\,, & z>0\,, \cr
                     -\ft{16}{25} k\,, & z<0\,, } \right.\label{kgrel10}
\ee
where $k$ is assumed to be positive.  This means that one cannot
strictly speaking view the domain wall as a solution within the massive
type IIA theory as formulated in \cite{romans10}, since there $m$ is a
fixed parameter in the Lagrangian.  However, the theory can be
re-expressed in a formulation where $m$ is replaced by a 10-form field,
with the mass parameter now arising as a constant of integration.  It
now makes sense for the parameter to be only piecewise constant.  In
what follows, we shall implicitly assume that we are working with this
reformulation of the theory, which allows (\ref{kgrel10}) to hold.

     Note that the nine-dimensional flat Minkowskian spacetime $dx^\mu
dx_\mu$ of the solution (\ref{d8brane}) can be replaced by any
Ricci-flat Minkowski-signatured spacetime \cite{prt}.  On other hand, it was
observed that domain walls associated with D$p$-branes with $p\ge 6$
cannot trap gravity.  Nevertheless, one can still obtain gravity on the
world-volume in such a case by locating the branes at orbifold points,
so that the spacetime is compact \cite{losw}.  In this case, we would
expect that the resulting theory on the world-volume of the D8-brane
would be the ungauged $N=1$, $D=9$ supergravity.  We shall prove in the
next subsection that this can indeed be obtained from the massive type
IIA theory via a consistent brane-world Kaluza-Klein reduction.

\subsection{$N=1$ supergravity in $D=9$ from massive type IIA}

    We find that the following Kaluza-Klein Ansatz for the
ten-dimensional massive type IIA fields yields a consistent reduction to
nine dimensions:
\bea
&&d\hat s_{10}^2 = e^{-\fft{5}{16} \sqrt{\fft27} \phi}\, 
W^{\fft2{25}}\, ds_9^2 +
e^{\fft{35}{16}\sqrt{\fft27}\phi}\, dz^2\,,\nn\\
&&\hat A_\1=0\,,\qquad 
\hat A_\2= \fft1{2m}\, W^{\fft{16}{25}}\, F_\2\,,\qquad
\hat A_\3 = \fft1{4m}\, W^{\fft{32}{25}}\, F_\3\,,\label{10to9ans}\\
&&e^{\hat\phi} = W^{-\fft45}\, e^{-\fft78\sqrt{\fft27}\phi}\,,\nn
\eea
where $W$ is given by (\ref{wdef}) and $g$ is related to $k$ by
(\ref{kgrel10}).  Substituting this Ansatz into the massive type IIA
equations of motion, we find that they are all satisfied provided that
the nine-dimensional fields $ds_9^2$, $\varphi$, $F_\2=dA_\1$ and
$F_\3=dA_\2 -\ft12 A_\1\wedge F_\2$ satisfy the equations of motion of
nine-dimensional ungauged simple supergravity.  These equations can be
derived from the Lagrangian
\be
{\cal L}_9 = R\, {*\oneone} - \ft12 {*d\phi}\wedge d\phi -
\ft12 e^{-\sqrt{\fft87}\phi}\, {*F_\3}\wedge F_\3 - \ft12 
e^{-\sqrt{\fft27}\phi}\, {*F_\2}\wedge F_\2\,.
\ee

\subsection{Branes on the D8-brane}

        Having consistently embedded the ungauged $N=1$, $D=9$
supergravity in massive type IIA supergravity, we can lift all the
solutions of this nine-dimensional theory back to $D=10$.  The
nine-dimensional theory supports BPS $p$-branes such as the string,
4-brane, black hole and 5-brane.  These solutions are straightforward
and well known.  When they are lifted back to massive type IIA
supergravity using the Ansatz (\ref{10to9ans}), they can be viewed as
branes living on the D8-brane.

\bigskip\bigskip
\noindent{\it Case 1: String on D8-brane:}
\bigskip

The solution of the $D=9$ string lifted back to $D=10$ becomes
\bea
d\hat s_{10}^2 &=& W^{\ft2{25}}\Big[H^{-\ft58} (-dt^2 + dx^2) +
H^{\ft38} (dr^2 + r^2 d\Omega_6^2)\Big] + H^{-\ft58} dz^2\,,\nn\\
\hat A_\3 &=& \fft{1}{4m}\, W^{\ft{32}{25}}\, dt\wedge dx\wedge
dH^{-1}\,,\nn\\
e^{\hat \phi} &=& W^{-\ft45} \, H^{\ft14}\,,\qquad
H=1 + \fft{Q}{r^5}\,.\label{1on8}
\eea
This solution can be viewed as a D2-brane ending on the D8-brane, with
the end points forming a string.  To see this, it is helpful to
introduce a new coordinate $y$ in place of $z$, defined by $dy =
\ft{24}{25}\, W^{-\fft1{25}}\, dz$, and hence
\be
W= (1+ k\, |y|)^{\fft{25}{24}}\,.\label{ztoy}
\ee
Using this variable, the $y$-dependence of the metric is extracted as an
overall conformal factor, and we have
\be
d\hat s_{10}^2 = (1 + k|y|)^{\ft1{12}}\Big[
H^{-\ft58}(-dt^2 + dx^2 + dy^2) + H^{\ft38}(dr^2 + r^2 d\Omega_6^2)
\Big]\,.\label{2endon8}
\ee

\bigskip\bigskip
\noindent{\it Case 2: 4-brane on D8-brane:}
\bigskip

     The $D=9$ 4-brane solution (the magnetic dual of the string
solution), lifted back to $D=10$, becomes
\bea
d\hat s_{10}^2 &=& W^{\ft2{25}}\Big[H^{-\ft38}\, dx^\mu dx_\mu +
H^{\ft58}(dr^2 + r^2 d\Omega_3^2)\Big] + H^{\ft58} dz^2\,,\nn\\
\hat A_\3 &=& \fft{Q}{4m} W^{\ft{32}{25}}\, \Omega_3\,,\nn\\
e^{\hat \phi} &=& W^{-\ft45}\, H^{-\ft14}\,,\qquad
H=1 + \fft{Q}{r^2}\,.\label{4on8}
\eea
Using the same coordinate transformation (\ref{ztoy}),  
the metric can be re-expressed as 
\be
d\hat s_{10}^2 = (1 + k|y|)^{\ft1{12}}\Big[
H^{-\ft38}\,dx^\mu dx_\mu + H^{\ft58}(dy^2+dr^2 + r^2 d\Omega_3^2)
\Big]\,.\label{4inton8}
\ee
Thus the solution can be viewed as a D4-brane intersecting with a
D8-brane, with the D4-brane uniformly delocalised on the 1-dimensional
transverse space of the D8-brane.

       In Case 1 above, the intersection of the D2-brane and the
D8-brane is such that the overall world-volume is a string, and the
solution describes a D2-brane ending on the D8-brane.  In Case 2, the
intersection of the D4-brane and D8-brane is such that the overall
world-volume is the entire 4-brane, and so the solution describes a
D4-brane living in the D8-brane.

\bigskip\bigskip
\noindent{\it Case 3: Black hole on D8-brane:}
\bigskip

       The black hole solution of the $D=9$ theory can be lifted to
$D=10$, where it becomes
\bea
d\hat s^2_{10} &=& W^{\ft2{25}} \Big[-H^{-\ft{13}{8}}\, dt^2
+ H^{\ft38}(dr^2 + r^2 d\Omega_7^2)\Big] + H^{-\ft58}\, dz^2\,,\nn\\
\hat A_\2&=&\fft1{2m} W^{\ft{16}{25}}\, dt\wedge dH^{-1}\,,\nn\\
e^{\hat \phi} &=& W^{-\fft45} H^{\ft14}\,,\qquad
H=1 + \fft{Q}{r^6}\,.\label{0on8}
\eea
Using (\ref{ztoy}), the metric can be cast into the
form
\be
d\hat s_{10}^2 = (1 + k|y|)^{\ft1{12}}\Big[-H^{-\ft{13}8}\,dt^2 +
H^{-\ft58}\, dy^2 +  H^{\ft38}(dr^2 + r^2 d\Omega_7^2)
\Big]\,.\label{01on8}
\ee
The solution can be viewed as the intersection of an NS-NS string and a
D0-brane with the D8-brane. In particular, the string NS-NS string ends
on the D8-brane whilst the D0-brane lives in the D8-brane.  To see this,
we note that a standard solution for the intersection of a string and a
D0-brane would be
\be
ds_{10}^2 = -H_0^{-\fft78}\, H_1^{-\fft34}\, dt^2 + H_0^{\fft18}\,
H_1^{-\fft34}\, du^2 + H_0^{\fft18}\, H_1^{\fft14}\, d\vec y^2\,,
\ee
where $H_0$ and $H_1$ are independent harmonic functions on the
eight-dimensional common transverse space of the $\vec y$ coordinates.
If these two harmonic functions are set equal, $H_0=H_1=H$, then we
obtain the structure found in (\ref{01on8}).

\bigskip\bigskip
\noindent{\it Case 4: 5-brane on D8-brane:}
\bigskip

      The 5-brane solution (the magnetic dual of the black hole) can be
lifted to $D=10$, where it becomes
\bea
d\hat s^2_{10} &=& W^{\ft2{25}} \Big[ H^{-\ft38}\, dx^\mu dx_\mu
+ H^{\ft{13}8} (dr^2 + r^2 d\Omega_2)\Big] + H^{\ft58}\, dz^2\,,\nn\\
\hat A_\2&=&\fft{Q}{2m} W^{\ft{16}{25}}\, \Omega_2 \,,\nn\\
e^{\hat \phi} &=& W^{-\fft45}\, H^{-\ft14}\,,\qquad
H=1 + \fft{Q}{r}\,.\label{5on8}
\eea
Using the redefinition (\ref{ztoy}), the metric can be cast into the 
form
\be
d\hat s_{10}^2 = (1 + k|y|)^{\ft1{12}}\Big(-H^{-\ft38}\,dx^\mu dx_\mu +
H^{\ft58}\, dy^2 +  H^{\ft{13}8}(dr^2 + r^2 d\Omega_2^2)
\Big)\,.\label{xxxon8}
\ee
The solution can be viewed as an intersection of an NS-NS 5-brane and a
D6-brane with the D8-brane.  In particular, the NS-NS 5-brane lives in
the D8-brane, whilst the D6-brane ends on the D8-brane.  (The standard
solution for the intersection of a 5-brane and a 6-brane would be of the
form
\be
ds_{10}^2 = H_5^{-\fft14}\, H_6^{-\fft18}\, dx^\mu \, dx_\mu +
H_5^{\fft34}\, H_6^{-\fft18}\, du^2 + H_5^{\fft34}\, H_6^{\fft78}\,
d\vec y^3\,,
\ee
where $H_5$ and $H_6$ are independent harmonic functions on the common
transverse 3-space of the $\vec y$ coordinates.  In our case, the two
harmonic functions are equal, $H_5=H_6=H$.)

    Note that it is straightforward also to construct pp-wave and
Taub-NUT solutions on the world-volume of the D8-brane.

\section{Reductions of $SU(2)$-gauged $D=8$ supergravity}

\subsection{Brane-world reduction to $D=7$}

   Although there is no gauged supergravity in eight dimensions that
admits a maximally-symmetric AdS$_8$ solution, there {\it is} a gauged
theory that arises from the dimensional reduction of eleven-dimensional
supergravity on $S^3$ \cite{sase}.  Since only the gauge bosons of the
left-acting $SU(2)$ of the $SO(4)\sim SU(2)\times SU(2)$ are retained in
the truncation, the consistency of this reduction to $D=8$ is guaranteed
by the standard group-theoretic arguments of \cite{scsc}.  The theory
can also be obtained from the $S^2$ reduction of type IIA theory, and
the $SU(2)$ is the isometry group of the 2-sphere. The eleventh
coordinate is the fibre coordinate of the $S^3$, which can be viewed as
a $U(1)$ bundle over $S^2$ \cite{kksphere}.  The eight-dimensional
theory admits a dilatonic 6-brane domain-wall solution, and this
provides a starting-point for the construction of a brane-world
Kaluza-Klein reduction to $D=7$.

     The bosonic sector of the eight-dimensional theory contains the
metric, a dilatonic scalar $\varphi$, five further scalars that can be
parameterised by a unimodular $3\times 3$ symmetric matrix $T_{ij}$, the
$SU(2)$ Yang-Mills potentials $A_\1^i$, three 2-form potentials
$B_\2^i$, and a 3-form potential $A_\3$.  The description of the theory
is a little involved, but the majority of the complications come from
the scalars $T_{ij}$ and the Yang-Mills potentials $A_\1^i$ that will in
fact be set to zero in our brane-world Kaluza-Klein reduction to $D=7$.
It is not in general consistent in $D=8$ to set $T_{ij}=\delta_{ij}$ and
$A_\1^i=0$ while keeping all the other fields non-vanishing, since the
retained fields will act as sources for those that are set to zero.
However, since in our brane-world reduction to $D=7$ the Ansatz for the
remaining non-vanishing eight-dimensional fields will be such that these
source terms vanish, it is sufficient for our purposes to present the
truncated eight-dimensional theory, together with constraints that will
be identically satisfied by the brane-world reduction Ansatz.  These
constraints are precisely the conditions that the sources that would
have excited the truncated fields should be zero.

    It is in fact easy to obtain this truncation of the
eight-dimensional gauged theory as an $S^3$ reduction from $D=11$.  The
Ansatz is given by
\bea
d\hat s_{11}^2 &=& e^{-\fft13\varphi}\, ds_8^2 + e^{\fft23\varphi}\,
g^{-2}\, d\Omega_3^2\,,\nn\\
\hat A_\3 &=& A_3 + \ft12 g^{-1}\,  B_\2^i\wedge \sigma_i\,.\label{11to8ans}
\eea
The quantities $\sigma_i$ are the three left-invariant 1-forms on the
group manifold $SU(2)$, satisfying $d\sigma_i = -\ft12 \ep_{ijk}\,
\sigma_j\wedge \sigma_k$.  In terms of these, the unit metric on $S^3$
can be written as $d\Omega_3^2 = \ft14 \sigma_i\, \sigma_i$.
Substituting the Ansatz into the bosonic equations of motion of
eleven-dimensional supergravity,
\be
d\hat F_\4 = \ft12 \hat F_\4\wedge \hat F_\4\,,\qquad
\hat R_{MN} = \ft1{12} (\hat F^2_{MN} - \ft1{12} \hat F_\4^2\, \hat
g_{MN})\,,\label{d11eom}
\ee
we find that the field equation for $\hat F_\4$ implies
\bea
d(e^{\varphi}\, {*F_\4}) &=& - 2g\, B_\2^i\wedge G_\3^i\,,\nn\\
d{*G_\3^i} &=& -4g^2\, e^{-\varphi}\, {*B_\2^i} - 2g\, F_\4\wedge
B_\2^i - g\, \ep_{ijk}\, G_\3^j\wedge G_\3^g\,,\nn\\
F_\4 \wedge F_\4 &=&0\,,\label{f4eom8}
\eea
where $F_\4\equiv dA_\3$ and $G_\3^i \equiv d B_\2^i$.  Note that the
last equation in (\ref{f4eom8}) is one of the constraints that results
from our having truncated out the $T_{ij}$ and $A_\1^i$ fields.  From
the Einstein equation in (\ref{d11eom}), we obtain the following
eight-dimensional equations of motion,
\bea
R_{\mu\nu} &=& \ft12 \del_\mu\varphi\, \del_\nu\varphi - \ft16
\square\varphi\, g_{\mu\nu} + \ft1{12} e^{\varphi}\,
[F_{\mu\rho\sigma\lambda}\, F_\nu{}^{\rho\sigma\lambda} 
- \ft1{12} F_\4^2\, g_{\mu\nu} ]\nn\\
&&+ \ft14 [G^i_{\mu\rho\sigma}\, G_\nu^{i\, \rho\sigma} - \ft19
(G_\3^i)^2\, g_{\mu\nu} ] 
+ 2g^2\, e^{-\varphi}\, [ B^i_{\mu\rho}\, B_\nu^{i\, \rho} - \ft16
(B_\2^i)^2\, g_{\mu\nu}]\,,\nn\\
\square\varphi &=& 6 g^2\, e^{-\varphi} + \ft1{48} e^{\varphi}\,
F_\4^2 - g^2 e^{-\varphi}\, (B_\2^i)^2\,,\label{einst8}
\eea
together with the further constraints
\bea
e^{\varphi}\, F_{\mu\nu\rho\sigma}\, G^{i\, \nu\rho\sigma} + 6g\,
\ep_{ijk}\, G^j_{\mu\rho\sigma}\, B^{k\, \rho\sigma} &=&0\,,\nn\\
e^{\varphi}\, G^i_{\mu\nu\rho}\, G^{j\, \mu\nu\rho} - 12 g^2\,
B^i_{\mu\nu}\, B^{j\, \mu\nu} &=&0\,.\label{con8}
\eea
These come from the mixed and the purely internal components of the
eleven-dimensional Einstein equation respectively.

    The eight-dimensional equations of motion admit a domain-wall
``ground-state'' solution, where all fields except $ds_8^2$ and
$\varphi$ are set to zero, and
\be
ds_8^2 = W^{\fft23}\, dx\cdot dx + dz^2\,,\qquad
e^{\varphi} = W^2\,,
\ee
where 
\be
W = 1 + k\, |z|\,,\qquad k^2 = \ft94 g^2\,.
\ee
Specifically, 
\be
g=  \left\{\matrix{ \ft23 k\,, & z>0\,, \cr
                    -\ft23 k\,, & z<0\,. } \right. \label{kgrel8}
\ee
(As usual, $g$ is allowed to have the necessary sign-change across the
domain-wall provided that one thinks of obtaining the eight-dimensional
gauged theory as an $S^3$ reduction from $D=11$, since then $g$ arises
as a constant of integration, rather than as a fixed parameter in the
eight-dimensional Lagrangian.)

   This motivates the construction of the following brane-world
reduction Ansatz, to give ungauged seven-dimensional $N=2$ supergravity
from the gauged eight-dimensional theory:
\bea
d\hat s_8^2 &=& e^{-\fft1{2\sqrt5}\, \phi}\, W^{\fft23}\, ds_7^2 + 
e^{\fft{\sqrt5}{2}\phi}\, dz^2\,,\nn\\
\hat B_\2^i &=& \ft1{2\sqrt2}\, g^{-1}\, W^{\fft43}\, F_\2^i\,,\nn\\
\hat A_\3 &=& A_\3\,,\qquad e^{\hat \varphi} = W^2\,
e^{\fft{\sqrt5}{2}\phi}\,,\label{8to7ans}
\eea
where we have now placed hats on all the eight-dimensional fields.

     Substituting this Ansatz into the equations of motion for the
eight-dimensional gauged theory given above, we find that they are
satisfied provided that the seven-dimensional fields satisfy the
equations of motion of ungauged seven-dimensional $N=2$ supergravity.
Specifically, these can be derived from the Lagrangian
\be
{\cal L}_7 = R\, {*\oneone} - \ft12 {*d\phi}\wedge d\phi - \ft12
e^{\sqrt{\fft85} \phi}\, {*F_\4}\wedge F_\4 - \ft12 e^{-\sqrt{\fft25}\phi}\,
{*F_\2^i}\wedge F_\2^i - \ft12 F_\2^i\wedge F_\2^i\wedge A_\3\,.
\ee
It should be noted also that the Ansatz (\ref{8to7ans}) identically
satisfies the constraint equations in (\ref{f4eom8}) and (\ref{con8}),
and so indeed our assumption that these would eventually be satisfied in
the brane-world reduction is justified.  Note that the theory naturally
arises with a 4-form field strength rather than the 3-form field
strength that would naturally come from the $T^3$ reduction of the
heterotic theory.  This suggests that the former and the latter can be
related by a strong/weak duality.

   Having obtained the brane-world reduction from eight-dimensional
gauged supergravity, we may now lift it back to $D=11$, by using the
$S^3$ reduction Ansatz (\ref{11to8ans}).  Thus we find that the
eleven-dimensional fields are given in terms of seven-dimensional fields
by
\bea
d\hat s_{11}^2 &=& e^{-\sqrt{\fft{8}{45}}\phi}\, ds_7^2 + 
e^{\sqrt{\fft5{18}}\phi}\, \Big[ W^{-\fft23}\, dz^2 + g^{-2}\,
W^{\fft43}\, d\Omega_3^2\Big]\,,\nn\\
\hat A_\3 &=& A_\3 + \ft1{4\sqrt2}\, g^{-2}\, W^{\fft43}\,
F_\2^i\wedge \sigma_i\,.\label{11to7ans}
\eea

     It is interesting to note that if we perform a coordinate
transformation from $z$ to $r$, defined by $W^{-\fft13}\, dz =dr$, and
hence
\be
W= (g\, r)^{\fft32}\,,
\ee
then this Ansatz for the reduction from $D=11$ to $D=7$ becomes 
\bea
d\hat s_{11}^2 &=& e^{-\sqrt{\fft{8}{45}}\phi}\, 
ds_7^2 +  e^{\sqrt{\fft5{18}}\phi}\,
(dr^2 + r^2\, d\Omega_3^2)\,,\nn\\
\hat A_\3 &=& A_\3 + \ft1{4\sqrt2}\, r^2\, F_\2^i\wedge\sigma_i\,.
\label{11to7ans2}
\eea
This is recognisable as a standard type of Kaluza-Klein reduction on
$T^4$, in which a truncation to the fields of $N=2$ supergravity in
$D=7$ has been performed.

\subsection{Brane-world reduction to $D=7$, from type IIA supergravity}

   Taking the results of the previous subsection, we can perform an
additional $S^1$ Kaluza-Klein reduction on the Hopf fibres of the
compactifying 3-sphere that was used in the reduction from $D=11$ to
$D=8$, thereby allowing us to obtain a brane-world reduction to $D=7$
that can be viewed as coming from type IIA supergravity compactified
first on $S^2$.

   To implement this procedure, we first specialise some results for the
Hopf reduction of $S^3$ that were obtained in \cite{clpcpn}.  In terms
of Euler angles $(\theta,\varphi,\psi)$, the three left-invariant
1-forms of $SU(2)$ can be written as
\bea
\sigma_1 &=& \cos\psi\, d\theta + \sin\psi\, \sin\theta\,
d\varphi\,,\nn\\
\sigma_2 &=& -\sin\psi\, d\theta +  \cos\psi\, \sin\theta\,
d\varphi\,,\nn\\
\sigma_3 &=& d\psi + \cos\theta\, d\varphi\,.
\eea
Clearly $\del/\del\varphi$ is a Killing vector for the 3-sphere metric
$d\Omega_3^2 = \ft14\sigma_i\, \sigma_i$, and it also leaves the 3-form
Ansatz in (\ref{11to7ans}) invariant.  Let $\mu_i$ be three coordinates
on $\R^3$ subject to the constraint $\mu_i\, \mu_i=1$, given in terms of
$\theta$ and $\psi$ by
\be
\mu_1=\sin\theta\, \sin\psi\,,\qquad \mu_2= \sin\theta\,
\cos\psi\,,\qquad \mu_3 = \cos\theta\,.
\ee
It is easily seen that in terms of these we can write the
left-invariant 1-forms as
\be
\sigma_i = - \ep_{ijk}\, \mu_j\, d\mu_k + \mu_i\, (d\varphi +
\cos\theta\, d\psi)\,.\label{limu}
\ee

   We can now perform a Kaluza-Klein $S^1$ reduction of the
eleven-dimensional expressions (\ref{11to7ans}) on the Hopf fibre
coordinate $\varphi$, using the standard Ansatz
\bea
d\hat s_{11}^2 &=& e^{-\fft16 \Phi}\, d\bar s_{10}^2 + e^{\fft43\Phi}\,
(d\varphi+ \bar \cA_\1)^2\,,\nn\\
\hat A_\3 &=& \bar A_\3 + \bar A_\2\wedge (d\varphi + \bar\cA_\1)\,,
\eea
where $\Phi$ is the type IIA dilaton.  Using (\ref{limu}), we therefore
obtain the following reduction Ansatz for the fields of type IIA
supergravity:
\bea
d\bar s_{10}^2 &=& W^{\fft16}\, \Big[ e^{-\fft{9}{8\sqrt{10}}\,
\phi} \, ds_7^2 + W^{-\fft23}\, e^{\fft{15}{8\sqrt{10}}\, \phi}\, dz^2 
+ \ft14 g^{-2}\, W^{\fft43}\, 
 e^{\fft{15}{8\sqrt{10}}\, \phi}\, d\Omega_2^2\Big]\,,\nn\\
\bar A_\3 &=& A_\3 + \ft1{4\sqrt2}\, g^{-2}\, W^{\fft43}\, 
\ep_{ijk}\, \mu_i\, F_\2^j\wedge d\mu_k\,,\nn\\
\bar A_\2 &=& \ft1{2\sqrt2}\, g^{-1}\, W^{\fft43}\, \mu_i\, F_\2^i\,,\nn\\
\bar \cA_\1 &=& \ft12 g^{-1}\, \cos\theta\, d\psi\,,\nn\\
e^{\Phi} &=& W\, e^{\fft{5}{4\sqrt{10}}\, \phi}\,,
\label{2as2red}
\eea
where the unit 2-sphere metric $d\Omega_2^2$ is given by
\be
d\Omega_2^2 = d\mu_i\, d\mu_i = d\theta^2 + \sin^2\theta\, d\psi^2\,.
\ee
The ``vacuum'' solution, corresponding to the metric (\ref{2as2red})
with $\phi=0$, can be viewed as the near-horizon limit of a D6-brane.

\subsection{Branes on the D6-brane}

     The $D=7$, $N=2$ supergravity admits membrane and string solutions
supported by electric or magnetic charges for $F_\4$.  When lifted back
to $D=11$, the membrane becomes an M2-brane delocalised on a
4-hyperplane, whilst the string can be viewed as an M5-brane wrapped on
the 4-hyperplane.  The seven-dimensional theory also admits black hole
and 3-brane solutions, which can be viewed as intersections of two
M2-branes, and intersections of two M5-branes, respectively.  From the
type IIA point of view, they can be viewed as membranes, strings, black
holes or 3-branes living in a D6-brane.

\section{Reductions of gauged maximal $D=7$ supergravity}

\subsection{Gauged maximal seven-dimensional supergravity}

   The bosonic Lagrangian for maximal $SO(5)$-gauged supergravity in 
$D=7$ can be written as 
\bea
{\cal L}_7 &=& \hat R\, {\hat *\oneone} -
\ft14 T^{-1}_{ij}\, {*D T_{jk}}\wedge
T^{-1}_{k\ell}\, D T_{\ell i}
- -\ft1{4}\, T^{-1}_{ik}\, T^{-1}_{j\ell}\, {\hat * \hat F_\2^{ij}}
\wedge \hat F_\2^{k\ell}
- -\ft12 T_{ij}\, {\hat *\hat S_\3^i}\wedge \hat S_\3^j \nn\\
&&+ \fft1{2g} \hat S_\3^i\wedge \hat H_\4^i -
\fft1{8g}  \ep_{i j_1\cdots j_4}\, \hat S_\3^i\wedge \hat F_\2^{j_1 j_2}\wedge
\hat F_\2^{j_3 j_4} + \fr1g \Omega_\7 - V\, {\hat *\oneone}\,,\label{d7lag}
\eea
where
\be
\hat H_\4^i \equiv D \hat S_\3^i = d\hat S_\3^i + 
g\, \hat A_\1^{ij}\wedge \hat S_\3^j\,.
\label{h4def}
\ee
The potential $V$ is given by
\be
V = \ft12  g^2 \Big(2 T_{ij}\, T_{ij} - (T_{ii})^2 \Big)\,,
\ee
and $\Omega_\7$ is a Chern-Simons type of term built from the
Yang-Mills fields, which has the property that its variation with
respect to $\hat A_\1^{ij}$ gives
\be
\delta \Omega_\7 =
\ft34 \delta_{i_1 i_2 k\ell}^{j_1 j_2 j_3 j_4}\, \hat F_\2^{i_1 i_2}\wedge
\hat F_\2^{j_1 j_2}\wedge  \hat F_\2^{j_3 j_4}\wedge 
\delta \hat A_\1^{k\ell}\,.
\ee
An explicit expression for $\Omega_\7$ can be found in \cite{ppn}.  Note
that the $S_\3^i$ are viewed as fundamental fields in the Lagrangian.
The symmetric unimodular $SO(5)$-valued tensor $T_{ij}$ describes the 14
scalar fields.

    Let us now set the $SO(5)$ Yang-Mills potentials $A_\1^{ij}$ to
zero, and take the scalars to be trivial also, $T_{ij}=\delta_{ij}$.
This is not in general a consistent truncation, since the remaining
fields $\hat S_\3^i$ would act as sources for the Yang-Mills and scalar
fields that have been set to zero. If we impose that these source terms
vanish, \ie
\be
\hat S_\3^i\wedge \hat S_\3^j=0\,,\qquad {\hat *\hat S_\3^i}\wedge
\hat S_\3^j=0\,,\label{s3sources}
\ee
then the truncation will be consistent.  (As we shall see below, these
sources terms will indeed vanish in the brane-world reduction that we
shall be considering.)  The remaining equations of
motion following from (\ref{d7lag}) are then
\bea
&&d{\hat *\hat S_\3^i} = 0\,,\qquad d\hat S_\3^i = g\, {\hat *\hat
S_\3^i}\,,\nn\\
&& \hat R_{AB} = \ft14 (\hat S^i_{ACD}\, \hat S^i_B{}^{CD} -
\ft2{15}\, (S^i_\3)^2\, \hat g_{AB}) - \ft32 g^2\, \hat g_{AB}\,.
\eea

\subsection{Chiral $N=(2,0)$ supergravity from $D=7$}

    We find that the following Kaluza-Klein Ansatz for the seven-dimensional
fields yields a consistent reduction to six dimensions:
\bea
&&d\hat s_7^2 = e^{-2k\, |z|}\, ds_6^2 + dz^2\,,\nn\\
&&\hat S_\3^i = e^{-2k\, |z|}\, F_\3^i\,,\qquad \hat
A_\1^{ij}=0\,,\qquad T_{ij} = \delta_{ij}\,,\label{d7ans}
\eea
where the constant $k$ is related to the gauge coupling constant $g$
by
\be
g=  \left\{\matrix{ -2k\,, & z>0\,, \cr
                    +2k\,, & z<0\,. } \right. \label{kgrel}
\ee
Substituting this Ansatz into the field equations of seven-dimensional
$SO(5)$-gauged supergravity, we find that all the equations are
consistently satisfied provided that the six-dimensional fields $ds_6^2$
and $F_\3^i$ satisfy the equations of motion of six-dimensional ungauged
$N=(2,0)$ chiral supergravity, namely
\be 
F_\3^i ={*F_3^i}\,,\qquad dF_\3^i=0\,,\qquad R_{\mu\nu} = \ft14
F^i_{\mu\rho\sigma}\, F^i_\nu{}^{\rho\sigma}\,.  \label{d6eom}
\ee 
Note that the self-duality of the 3-forms ensures that the constraints
(\ref{s3sources}) are indeed satisfied, since $F_\3^i\wedge F_\3^j=0$
for any pair of self-dual 3-forms.  Of course the self-duality of the
$F_\3^i$ fields also implies one cannot write a covariant Lagrangian for
this theory.

   Since we know the exact embedding of seven-dimensional maximal
$SO(5)$-gauged supergravity in $D=11$, via the $S^4$ reduction, we can
lift the above Ansatz to an embedding in eleven-dimensional
supergravity.  Using the $S^4$ reduction Ansatz, we therefore obtain
\bea
d\hat s_{11}^2 &=& e^{-2k\, |z|}\, ds_6^2 + dz^2 + g^{-2}\, d\mu_i\,
d\mu_i\,,\nn\\
\hat F_\4 &=& \fft1{8g^3}\,  \ep_{i_1\cdots i_5}\,\mu_{i_1}\,
d\mu_{i_2}\wedge \cdots \wedge d\mu_{i_5} - g^{-1}\, d(\mu_i\,
e^{-2k\, |z|}\, F_\3^i)\,,\label{d11ansatz}
\eea
where $\mu_i$ are coordinates on $\R^5$, subject to the constraint
\be
\mu_i\, \mu_i =1\,,
\ee
which defines the unit 4-sphere.

\subsection{Five-dimensional $N=4$ ungauged supergravity from
$SO(5)$-gauged $D=6$ supergravity} 

   By dimensionally reducing the embedding (\ref{d11ansatz}) of
six-dimensional chiral $N=(2,0)$ supergravity on a circle in the
six-dimensional spacetime, we can obtain an embedding of
five-dimensional $N=4$ supergravity in type IIA supergravity.  Thus we
begin by performing a standard $S^1$ Kaluza-Klein reduction of the
six-dimensional fields,
\bea
ds_6^2 &=& e^{-2\a\, \phi}\, d\td s_5^2 + e^{6\a\, \phi}\,
(dx_5+\cA_\1)^2\,,\nn\\
F_\3^i &=& e^{-4\a\, \phi}\, {\td * \wtd F_\2^i} + \wtd
F_\2^i\wedge (dx_5 + \cA_\1)\,,\label{d6d5red}
\eea
where $\a=1/(2\sqrt6)$ and $\wtd F_\2^i=d\wtd A_\1^i$.  (Note that
the form of the reduction Ansatz for the six-dimensional fields
$F_\3^i$ is dictated by the fact that they are self dual.)  The theory
that results from this dimensional reduction is ungauged $N=4$
supergravity in $D=5$.  It is straightforward to show that the
equations of motion in $D=5$ that follow from substituting
(\ref{d6d5red}) into (\ref{d6eom}) are derivable from the Lagrangian
\be
{\cal L}_5 = \wtd R\, {\td *\oneone} - \ft12 {\td * d\phi}\wedge
d\phi - \ft12 e^{-4\a\, \phi}\, {\td *\wtd F_\2^i}\wedge \wtd
F_\2^i - \ft12 e^{8\a\, \phi}\, {\td * \cF_\2}\wedge \cF_\2 - \ft12
\wtd F_\2^i\wedge \wtd F_\2^i\wedge \cA_\1\,,\label{d5n4lag}
\ee
where $\cF_\2=d\cA_\1$.

   We now substitute (\ref{d6d5red}) into (\ref{d11ansatz}), and compare
it with a standard Kaluza-Klein $S^1$ reduction from $D=11$ to $D=10$:
\bea
d\hat s_{11}^2 &=& e^{-\fft16\Phi}\, ds_{10}^2 + e^{\fft43\Phi}\,
(dx_5 + A_\1)^2\,,\nn\\
\hat F_\4 &=& F_\4 + F_\3\wedge (dx_5 + A_\1)\,.
\eea
By doing this, we arrive at the Ansatz for the embedding of the
five-dimensional ungauged $N=4$ supergravity in type IIA supergravity:
\bea
ds_{10}^2 &=& e^{-\fft94 k\, |z| -\fft54\a\, \phi}\, d\td s_5^2 + 
e^{-\fft14 k\, |z| + \fft34 \a\, \phi}\, dz^2 + g^{-2}\, e^{-\fft14
k\, |z| + \fft34\a\, \phi}\, d\mu_i\, d\mu_i\,,\nn\\
F_\4 &=& \fft1{8g^3}\, \  \ep_{i_1\cdots i_5}\,\mu_{i_1}\,
d\mu_{i_2}\wedge \cdots \wedge d\mu_{i_5} -g^{-1}\,
e^{-4\a\,\phi}\, d(\mu_i\, e^{-2k\, |z|} {\td *\wtd
F_\2^i})\,,\nn\\
F_\3 &=& -g^{-1}\, d(\mu_i\, e^{-2k\, |z|}\, \wtd F_\2^i)\,,\nn\\
F_2 &=& \cF_\2\,,\qquad e^{\Phi} = e^{-\fft34 k\, |z|}\, e^{\fft92\a\,
\phi}\,.
\eea

\subsection{Chiral $N=(2,0)$ supergravity from type IIA NS5-brane}

    We showed in section 5.2 that the chiral $(2,0)$ six-dimensional
supergravity can be obtained as a brane-world Kaluza-Klein reduction
from maximal gauged supergravity in $D=7$, and in turn, this can be
obtained as an $S^4$ reduction from $D=11$.  It was shown recently that
one can take a singular limit of the $S^4$ reduction of
eleven-dimensional supergravity, in which the 4-sphere degenerates to
$S^3\times \R$ \cite{0005137}.  The reduction can then be reinterpreted
as an $S^3$ reduction of type IIA supergravity, yielding a maximal
$SO(4)$-gauged supergravity in $D=7$ that admits a domain-wall, but not
AdS$_7$, as a solution.  One may refer to this theory as a
``domain-wall'' supergravity.

    By applying this limiting procedure in the context of the
brane-world reduction to the $(2,0)$ supergravity in $D=6$ that we
constructed in sections 5.1 and 5.2, we can now obtain a brane-world
reduction of the $D=7$ domain-wall supergravity to the $(2,0)$ theory in
$D=6$.  Furthermore, we can lift this back, {\it via} its $S^3$
embedding, to an Ansatz for type IIA supergravity.  Rather than
repeating the details of how the singular limit is taken here, we shall
simply quote and make use of the general results already obtained in
\cite{0005137}.

    We begin by considering the brane-world reduction of the maximal
seven-dimensional $SO(4)$-gauged domain-wall supergravity.  As in our
previous examples, many of the fields are set to zero in the brane-world
reduction, and so rather than presenting the full seven-dimensional
theory obtained in \cite{0005137}, we shall instead give it in an
already-truncated form, where all but the participating fields have
already been set to zero.  As usual, we should add the cautionary remark
that one cannot in general consistently set these fields to zero while
allowing the remaining fields to take generic configurations.  But in
anticipation of the fact that the brane-world reduction {\it will} be
consistent, we can make the truncation provided that we take note also
of the consequent required constraints, which will be satisfied by the
brane-world reduction Ansatz.

   In this spirit, we therefore set to zero all the fields of the
seven-dimensional $SO(4)$-gauged domain-wall supergravity constructed in
\cite{0005137} except for the metric, the dilatonic scalar field $\phi$,
and the 3-forms $S_\3^0$ and $S_\3^\a$.  Note that $S_\3^0$ is viewed as
a 3-form field strength that is derived from a 2-form potential, whilst
the four 3-forms $S_\3^\a$ are viewed as independent fields in their own
right, which satisfy first-order equations of motion.  Defining
$\gamma=\sqrt{2/5}$, we read off from \cite{0005137} that the equations
of motion for these remaining fields will be
\bea
&&d{\hat *d\phi} = -\gamma\, e^{-2\gamma\, \phi}\, {\hat *
S_\3^0}\wedge S_\3^0 + \ft14\gamma\, e^{\fft12\gamma\, \phi}\, {\hat *
S_\3^\a}\wedge S_\3^\a - 4\gamma\, g^2\, e^{\gamma\, \phi}\, {\hat
*\oneone}\,,\nn\\
&&d(e^{-2\gamma\, \phi}\, {\hat * S_\3}) = 0\,,\qquad
dS_\3^0=0\,,\nn\\
&&dS_\3^\a = g\, e^{\fft12\gamma\, \phi}\, {\hat * S_\3^\a}\,,\nn\\
&&\hat R_{MN} = \ft12\del_M\phi\, \del_N\phi + \ft14 e^{-2\gamma\,
\phi}\, [S_{MPQ}^0\, S_N^{0\, PQ} - \ft2{15} (S_\3^0)^2\, \hat g_{MN}
]\nn\\
&&\qquad\qquad + \ft14 e^{\fft12\gamma\, \phi}\, 
[S_{MPQ}^\a\, S_N^{\a\, PQ} - \ft2{15} (S_\3^\a)^2\, \hat g_{MN} ]
 -\ft45 g^2\, e^{\gamma\, \phi}\, \hat g_{MN}\,.\label{d7dweom}
\eea
(Note that we have placed hats on all quantities associated with the
seven-dimensional metric, in anticipation of the upcoming brane-world
reduction to $D=6$.)

    We may first note that these equations of motion admit a
domain-wall solution given by $d\hat s_7^2 = W^2\, dx\cdot dx + dz^2$,
$e^{-\gamma\, \phi} = W^2$, where $W=1+ k\, |z|$ and $k^2=4 g^2/25$.
This provides the basis for the brane-world reduction Ansatz to
$D=6$.  Specifically, we find that all the equations given in
(\ref{d7dweom}) are satisfied if we make the following Ansatz,
\bea
&&d\hat s_7^2 = W^2\, ds_6^2 + dz^2\,,\qquad e^{-\gamma\, \phi} = W^2\,,
\nn\\
&&S_\3^0 = F_\3^0\,,\qquad S_\3^\a = W^{\fft52}\, F_\3^\a\,,
\label{7to6dwans}
\eea
where the fields $ds_7^2$ and $F_\3^i$ satisfy the equations of motion
of ungauged six-dimensional $(2,0)$ supergravity, as given in
(\ref{d6eom}).  Note that here, the index $i$ on $F_\3^i$ runs over
the five values $i=(0,\a)$.  

    We can also use the results in \cite{0005137} to
lift the seven-dimensional fields to those of ten-dimensional type IIA
supergravity.  For the truncated system that we are considering here, 
we find that the embedding is simply given by
\bea
d\bar s_{10}^2 &=& e^{\fft38\gamma\, \phi}\, ds_7^2 + g^{-2}\,
e^{-\fft58\, \gamma\, \phi}\, d\Omega_3^2\,,\nn\\
e^{\Phi} &=& e^{\fft54\gamma\, \phi}\,,\qquad \bar A_\1=0\,,\nn\\
\bar F_\4 &=& - e^{\fft12\gamma\, \phi}\, 
\mu_\a\, S_\3^\a + g^{-1}\, S_\3^\a\wedge d\mu_\a\,,\nn\\
\bar F_\3 &=& 2 g^{-3}\, \Omega_\3 + g^{-1}\, S_\3^0\,,
\eea
where $\mu_\a$ denote Cartesian coordinates on $\R^4$ subject to the
constraint $\mu_\a\, \mu_\a=1$ that defines the unit 3-sphere
with metric $d\Omega_3^2$ and volume form $\Omega_\3$.  The barred
fields are those of type IIA supergravity, with $\Phi$ being the type
IIA dilaton.

   Substituting our brane-world reduction Ansatz (\ref{7to6dwans})
into this, we obtain the following brane-world embedding of
six-dimensional $(2,0)$ supergravity in type IIA supergravity:
\bea
d\bar s_{10}^2 &=& W^{\fft54}\, (ds_6^2 + W^{-2}\, dz^2 + g^{-2}\,
d\Omega_\3^2)\,,\nn\\
e^{\Phi} &=& W^{-\fft52} \,,\qquad \bar A_\1=0\,,\nn\\
\bar F_\4 &=& - W^{\fft32}\, \mu_\a\, F_\3^\a + g^{-1}\, W^{\fft52}\, 
F_\3^\a\wedge d\mu_\a\,,\nn\\
\bar F_\3 &=& 2 g^{-3}\, \Omega_\3 + g^{-1}\, F_\3^0\,.
\eea

\subsection{Chiral $N=(1,0)$ supergravity from heterotic 5-brane}

    It was shown in \cite{cllp,chamsab,0005137} that one can obtain the
chiral $N=(1,0)$ theory from the $N=2$ $SU(2)$-gauged supergravity in
$D=7$ that admits an AdS$_7$ vacuum solution, {\it via} a brane-world
Kaluza-Klein reduction. There is also an $SU(2)$-gauged supergravity in
$D=7$ that admits a domain wall instead of AdS$_7$ as a vacuum solution.
This theory can be obtained from the $S^3$ reduction of $N=1$
supergravity in $D=10$, and its domain-wall solution is therefore the
$S^3$ reduction of the heterotic 5-brane.  Clearly it can also obtained
from the truncation of the $N=4$ $SO(4)$-gauged maximal supergravity
discussed in the previous subsection.  It is straightforward to reduce
the seven-dimensional theory or the heterotic theory in $D=10$ on the
world-volume of the 5-brane and obtain the chiral $N=(1,0)$
supergravity.  The reduction Ansatz is identical to that of section 5.4,
but with all the fields that carry the index $\a$ set to zero.

     Both the $(2,0)$ and $(1,0)$ theories admit a self-dual string
solution in $D=6$.  This solution can be lifted to $D=11$, where it
becomes a self-dual string living in the world-volume of M5-brane, which
can also be viewed as an open membrane ending on the M5-brane
\cite{bob}.  When lifted back to $D=10$ instead, it can be viewed as a
self-dual string living in the NS5-brane or the heterotic 5-brane.

\subsection{$N=(1,1)$ supergravity from $D=7$ gauged supergravity}

    So far, we obtained the chiral ungauged supergravity in $D=6$ from
gauged supergravity in $D=7$, which itself can be obtained from $S^4$
reduction of M-theory, or the $S^3$ reduction of the type IIA or
heterotic theories.  There also exist gauged supergravities in $D=7$
that give rise instead to the $N=(1,1)$ non-chiral theory in six
dimensions, through a brane-world Kaluza-Klein reduction.  One example is
the $SU(2)$-gauged supergravity that is the $S^1$ reduction of
eight-dimensional $SU(2)$-gauged supergravity, which itself can be
obtained from the $S^3$ reduction of eleven-dimensional supergravity
\cite{sase}, as we discussed in section 4.  This is because the
brane-world reduction of the eight-dimensional gauged supergravity gives
rise to $N=2$ supergravity in $D=7$.  If we perform a further $S^1$
reduction on a brane-world direction, the bulk gauged-supergravity in
$D=8$ becomes a gauged $N=4$ supergravity in $D=7$, whilst the
world-volume seven-dimensional ungauged theory becomes the $N=(1,1)$
ungauged theory in $D=6$.

       There should also be an $SO(4)$-gauged supergravity in $D=7$ that
gives rise to the $N=(1,1)$ theory in $D=6$.  This can be obtained from
the $S^3$ reduction of the type IIB theory.  The bulk T-duality of the
type IIA and type IIB theories then translates into a T-duality between
the $N=(1,1)$ and $N=(2,0)$ theories in the 5-brane world-volume.

\section{Conclusion}

   In this paper, we have constructed several new examples of
brane-world Kaluza-Klein reductions.  Our focus was to construct the
reductions with larger supersymmetry and in diverse dimensions that in
general involve consistent Kaluza-Klein reductions with dilatonic
co-dimension one objects, thus extending the results obtained in
\cite{bob} in several ways. Specifically, we have shown that it is
possible to construct consistent brane-world reductions of
five-dimensional $N=8$ $SO(6)$-gauged supergravity to ungauged $N=4$
supergravity in $D=4$; of massive type IIA supergravity to ungauged
$N=1$ supergravity in $D=9$; of eight-dimensional $N=2$ $SU(2)$-gauged
supergravity to ungauged $N=2$ supergravity in $D=7$; and of
seven-dimensional $N=4$ $SO(5)$-gauged supergravity to ungauged $N=2$
supergravity in $D=6$.  In all these cases, just as in the original
examples constructed in \cite{bob}, the degree of ungauged supersymmetry
in the lower dimension is one half of the gauged one in the higher
dimension, and in this paper we have focused mainly on the supergravity 
multiplets.

   A simple calculation shows that for any brane-world (co-dimension
one) reduction Ansatz of the form
\be
d\hat s^2 = e^{-2k\, |z|}\, ds^2 + dz^2\,,
\label{adsdw}
\ee
the Riemann tensor $\hat R_{ABCD}$ of the
$D$-dimensional metric $d\hat s^2$ satisfies
\be
\hat R_{ABCD}\, \hat R^{ABCD} = e^{4k\, |z|}\, R_{abcd}\, R^{abcd}\, -
4k^2\, e^{2k\, |z|}\, R + 2D(D-1)\, k^4\label{riem}
\ee
in the bulk, where $R_{abcd}$ and $R$ are the Riemann tensor and Ricci
scalar of the reduced metric $ds^2$.  This implies that any curvature of
the lower-dimensional metric for which $R_{abcd}\, R^{abcd}$ or $R$ is
non-zero, no matter how small, will lead to curvature singularities in
the higher-dimensional metric on the Cauchy horizons at $z=\pm\infty$.
These singularities were discussed in detail for a Schwarzschild black
hole on the brane in \cite{hawcha}.
In \cite{gidkatran}, it was argued that such curvature singularities on
the horizons arise as an artefact of considering only the zero-mode of
the metric tensor, and that if the massive Kaluza-Klein modes are taken
into account they could actually become dominant near the horizons, and
may lead to a finite curvature there.  The results of \cite{bob} and
this paper suggest that the phenomenon of diverging curvature on the
Cauchy horizons for the AdS domain wall reductions (or null horizons for
the dilatonic domain wall reductions) may be more severe.  Specifically
these results show that the brane-world reductions correspond to exact
fully non-linear consistent embeddings in which the massive Kaluza-Klein
modes can be consistently decoupled.  This implies that there certainly
exist exact solutions on the brane-world where Kaluza-Klein modes do not
enter the picture, even at the non-linear level.  For these solutions,
the curvature will inevitably diverge at the horizons.  It becomes
necessary, therefore, either to live with these (null) singularities or
else to find a principle, perhaps based on the imposition of appropriate
boundary conditions, for rejecting the solutions of this
type.\footnote{Let us also remark that a deviation of the dilatonic
domain wall solutions from the flat (BPS)-limit generically introduces
naked singularities \cite{cs}, again pointing towards difficulties with
the interpretation of such solutions within a more realistic set-up.}
  It should be emphasised, however, that regardless of the physical
questions that are prompted by these results, the brane-world
Kaluza-Klein reductions remain valid mathematical constructs in their
own right.  In fact as relatively simple examples of consistent
reductions that have no obvious group-theoretic explanation, they can
be viewed as precursors of the remarkable examples of consistent
reductions on spheres.

   Finally, we again emphasise that the absolute-value sign in
Eq. (\ref{adsdw}) for the brane-world metrics in the AdS co-dimension
one brane (as well as dilatonic examples as discussed in the text)
actually requires an explicit delta function source to support such a
$Z_2$-symmetric co-dimension one object that in turn allows for the
trapping of gravity on the world-volume of the brane (at $z=0$). The
understanding of such delta-function sources in the lower dimension
may require a subtle interpretation in terms of fundamental sources,
such as $D-$brane sources of the higher dimensional theory
\cite{cdlls}.  Nevertheless, the consistency of the Kaluza-Klein
reduction in the bulk (for $z\ne 0$) is valid quite independently of
the origin of the domain wall sources.

\section*{Acknowledgments} 

We should like to thank Gary Gibbons, Chris Hull, Jim Liu and 
Kelly Stelle for discussions.

\end{document}